\documentclass[aps,prl,singlecolumn,preprint,showpacs,superscriptaddress,groupedaddress]{revtex4-1}

\usepackage{graphicx}
\usepackage{verbatim}
\usepackage{float}
\usepackage{sidecap}
\usepackage{dcolumn}
\usepackage{bm}
\usepackage{amssymb}
\usepackage{amsmath}	
\usepackage{color}
\usepackage{multirow}
\usepackage{upgreek,xspace}
\usepackage[a4paper]{geometry}
\newgeometry{top=2cm,right=1.25cm,left=1.25cm,bottom=2cm}
\usepackage[colorlinks=true, urlcolor=blue, pdfborder={0 0 0}]{hyperref}
\hypersetup{
linkcolor=blue,
citecolor=blue
}

\begin{document}

\title{Understanding electric field control of electronic and optical properties of strongly-coupled multi-layer quantum dot molecules}

\author{Muhammad Usman} 
\email{usman@alumni.purdue.edu} 
\affiliation{School of Physics, The University of Melbourne, Parkville, 3010, Melbourne, VIC, Australia.}

\begin{abstract}
Strongly-coupled quantum dot molecules (QDMs) are widely deployed in the design of a variety of optoelectronic, photovoltaic, and quantum information devices. An efficient and optimized performance of these devices demands engineering of the electronic and optical properties of the underlying QDMs. The application of electric fields offers a knob to realise such control over the QDM characteristics for a desired device operation. We perform multi-million-atom atomistic tight-binding calculations to study the influence of electric fields on the electron and hole wave function confinements and symmetries, the ground-state transition energies, the band-gap wavelengths, and the optical transition modes. The electrical fields both parallel ($\vec{E_p}$) and anti-parallel ($\vec{E_a}$) to the growth direction are investigated to provide a comprehensive guide on the understanding of the electric field effects. The strain-induced asymmetry of the hybridized electron states is found to be weak and can be balanced by applying a small $\vec{E_a}$ electric field, of the order of 1 KV/cm. The strong interdot couplings completely break down at large electric fields, leading to single QD states confined at the opposite edges of the QDM. This mimics a transformation from a type-I band structure to a type-II band structure for the QDMs, which is a critical requirement for the design of intermediate-band solar cells (IBSC). The analysis of the field-dependent ground-state transition energies reveal that the QDM can be operated both as a high dipole moment device by applying large electric fields and as a high polarizibility device under the application of small electric field magnitudes. The quantum confined Stark effect (QCSE) red shifts the band-gap wavelength to 1.3 $\mu$m at the 15 KV/cm electric field; however the reduced electron-hole wave function overlaps lead to a decrease in the interband optical transition strengths by roughly three orders of magnitude. The study of the polarisation-resolved optical modes indicate the benefit of applying small electric fields, which lead to an isotropic polarisation response, a desirable property for the semiconductor optical amplifiers (SOAs).  
\end{abstract}


\maketitle

\subsection{1 $-$ Introduction}

Just as the quantum dots (QDs) are widely known as ``artificial atoms" due to their discrete energy spectra and the three-dimensional confinement of the charge carriers, the strongly-coupled quantum dots are often referred as ``artificial molecules" due to their hybridized bonding and anti-bonding type electronic states, closely resembling to that of the molecules made up of atoms. Self-assembled quantum dot molecules (QDMs) made up of III-V materials are one of the most popular types of nanostructures used in a variety of devices for optoelectronic~\cite{Takahashi_1, Alonso_1, Inoue_1, Usman_1, Usman_7, Usman_8, QDM_1}, photovoltaic~\cite{Kada_2015, Okada_2015, Oshima_1, Luque_1, Simmonds_1}, and quantum information technologies~\cite{XLi_2003, Chen_2015}. The QDMs typically grow in the form of vertical stacks due to the presence of strain, which stems from the lattice mismatch of the substrate and the QD material. One such example is of InAs/GaAs QDMs where about 7\% lattice mismatch between the InAs and GaAs materials results in the formation of nearly perfectly aligned, vertically stacked QD layers of InAs when epitaxially grown on top of a GaAs substrate~\cite{Xie_1, Inoue_SPIE_2010}. Similar stacking of the QD layers forming QDMs has been demonstrated in other material systems such as InAs/InGaAsP/InP~\cite{Roy_2010, Anantathanasarn_2006}, InGaAs/GaAs~\cite{Sugaya_2012, Kim_2013}, GaSb/GaAs~\cite{Peter_2012}, and InGaN/GaN~\cite{Sebald_2008, Schulz_2011}, etc. The stacking of multiple QD layers offers many useful benefits from the perspective of devices, such as increased number of photons emitted or absorbed per unit area at the QD energy levels for photo-diodes and lasers, a wider range of light absorption wavelengths for intermediate-band solar cells (IBSCs), access to long wavelengths for mid and far-infrared photo-detectors, and polarisation insensitivity of the gain for semiconductor optical amplifiers (SOA's), etc. The performance of these devices is critically related to the confinement of charge carriers inside the QDM. The spatial distributions and symmetries of the confined electron and hole wave functions control interband/intraband optical emissions and absorptions. Therefore it is of crucial significance to properly understand the electronic structure of the QDMs in order to fully exploit their promising properties for the design and optimization of devices.         

The electronic structure of self-assembled QDMs is strongly influenced by their geometry and the strain which stems from the lattice mismatch of the constituent materials. Therefore, the past theoretical studies of the QDMs have primarily focused on these two parameters to investigate the symmetries and the spatial confinements of the charge carriers. It has been experimentally shown that the strain not only plays a major role in governing inter-dot couplings leading to the formation of molecular states, but also strongly influences the polarisation response of the output photoluminescence (PL) spectra, leading to polarisation insensitivity of the output light for the properly designed QDMs~\cite{Inoue_1, Harada_1}. For many applications such as photodetectors, photovoltaics (such as IBSCs), and quantum information devices, the QDMs are also subject to electric fields, which can significantly influence their electronic structure. Such fields are either present internally such as investigated recently for the solar cells~\cite{Kasamatsu_1, Tomic_1}, or can be applied externally to purposely tune the properties of the devices for a certain operation~\cite{Chen_2015}. Although the impact of electric fields has been extensively studied in the literature for the single QDs~\cite{Jin_APL_2004, Fry_PRL_2000, Usman_IWCE_2009}, the bi-layer QDMs~\cite{Usman_4, Bester_PRB_2005, Skold_PRL_2013, QDM_1, Scheibner_PRB_2007, Szafran_PRB_2005}, and for the small QD stacks consisting of three to four QD layers~\cite{Szafran_PRB_2008}, to-date no theoretical understanding is available on the electronic and optical properties of the multi-layer QDMs under the influence of electrical fields. 

One rather intuitive effect of the electric fields is to Stark shift the energy levels of the QDMs and therefore modulate the emission/absorption peak wavelengths. However there are several open questions which require rigorous theoretical calculations, such as: (1) How the presence/application of electric fields modify the spatial confinements and symmetries of the electron and hole wave functions?, (2) What is the magnitude of the electric field required to break the coupling of quantum dot layers in a QDM?, (3) Does the impact of an applied electric field directional on the electronic and optical properties of the QDMs?, (4) What is the built-in dipole moment and polarizibility for the QDMs as a function of the electric field magnitudes and directions, and (5) What are interband optical transition strengths at large electric fields, where the band gap wavelength is red shifted to 1.3 $\mu$m. A detailed and reliable theoretical understanding of these unknown parameters is critically significant for the experimentalists working on the design and implementation of cutting-edge technologies based on QDMs. 

This work reports a first comprehensive theoretical study of the electronic and optical properties of the QDMs under the influence of electric fields. The physical insights provided are based on the state-of-the-art multi-million atom electronic structure and optical mode simulations, and deliver a reliable understanding of the complex physics of the QDMs, which involve subtle interplay between the applied electric fields and the strong interdot couplings. In order to answer the aforementioned outstanding questions, we apply electric fields along both parallel and anti-parallel directions to the growth direction of a QDM recently discussed in the experimental studies~\cite{Kada_2015, Okada_2015} and investigate its effects on the spatial confinements and symmetries of the electron and hole wave functions, band-gap wavelengths, ground state transition energies, and polarisation-resolved interband optical modes. Our results show that the application of a small electric field anti-parallel to the growth direction is sufficient balance the asymmetric effect of the underlying strain on the electron states. At large electric field magnitudes, the interdot couplings completely break down and lead to the formation of single QD electron and holes states, which are confined at the opposite ends of the QDM. This mimics a transformation of the QDM electronic structure from a type-I system to an effectively type-II system, in which the single QD electrons and holes are separated by a large distance, of the order of the size of the QDM along the growth direction -- a desirable property for the design of the IBSCs. The band-gap wavelength is calculated to increase roughly linearly as a function of the applied electric fields, irrespective of its direction for the large electric field magnitudes. However at the low electric fields ($<$ 2 KV/cm), the change in the band-gap wavelength is highly dependent on the direction of the electric field. An interesting property of the QDMs is revealed by analysing ground state transition energies and demonstrating that QDMs exhibit different characteristics under small and large fields: at the small fields, the QDMs have large polarizibilities but negligible dipole moments and this behaviour reverses at the large fields. The increased spatial separation between the electron and hole wave functions is directly reflected in the calculated interband optical transition strengths, which decrease as a function of the applied electric fields. At an electric field magnitude of 15 KV/cm required for the band-gap wavelength of 1.3 $\mu$m, the interband optical transition strength is calculated to decrease by nearly three orders of magnitude. Therefore very small oscillator gain is expected from the QDMs at the 1.3 $\mu$m operating wavelengths. However the application of small electric fields, ranging from 2 KV/cm to 5 KV/cm, are predicted to be useful for the optical devices demanding isotropic polarisation response such as semiconductor optical amplifiers.    

The rest of the paper is divided into the following sections: the section 2 briefly layouts the methods used in our simulations. The section 3 provides results and related discussions. We start with the section 3.1 by briefly reporting the electronic states of the QDM without the application of electric fields. This highlights the impact of strain on the electronic structure of the QDM and forms a basis for the remainder of the paper. The influence of the applied electric fields on the electron and hole wave functions is thoroughly investigated in the sections 3.2, 3.3, and 3.4, where we provide details about the field-dependent confinements and symmetries of the electronic states. The section 3.5 predicts the field-induced shifts in the band-gap wavelengths and the ground state transition energies. The section 3.6 provides calculations of the interband optical transitions for the TE and TM mode polarisations under the application of electric fields. Finally we conclude in the section 4.          

\subsection{2 $-$ Theoretical Framework}

Our investigation of the electric field effects on the QDM properties is based on multi-million-atom simulations of the strain, electronic structure and interband optical transitions. The strain is computed from the valence force field (VFF) relaxation of the atoms~\cite{Keating_1, Olga_1}. A very large simulation domain consisting of roughly 43 million atoms is resolved to properly accommodate the effects of the long-range strain fields and boundary effects. The electron/hole energies and states are obtained by solving a twenty-band $sp^3d^5s^*$ tight-binding Hamiltonian~\cite{Boykin_1}. The GaAs box surrounding the QDM have realistic boundary conditions as previously reported in Ref.~\onlinecite{Usman_5}, and the surface atoms are passivated according to the published recipe~\cite{Lee_1}. 

The polarisation-resolved interband optical transition strengths are calculated as follows: first we calculate interband optical transition strengths between the individual electron and hole states by using the Fermi's golden rule, where the squared absolute value of the momentum matrix elements is summed over the Kramer spin degenerate states: T$_{e_{i} \rightarrow h_{j}}$ = $|\langle e_{i}|[\overrightarrow{n},\mathbf{H}]|h_{j} \rangle |^{2}$ where $\mathbf{H}$ is the single particle tight binding Hamiltonian in the $sp^3d^5s^*$ basis, $e_{i}$ and $h_{j}$ are the $i$th and $j$th electron and hole states respectively, and $\overrightarrow{n}$ is a vector along the polarisation direction. The polarisation dependent optical modes TE$_{[110]}$, TE$_{[-110]}$, TE$_{[100]}$, TE$_{[010]}$, and TM$_{[001]}$ between the lowest conduction band state e$_1$ and the highest four valence band states h$_1$, h$_2$, h$_3$, and h$_4$ are calculated by rotating the polarisation vector $\overrightarrow{n}$ = $(\overrightarrow{x}+\overrightarrow{y}) \cos \phi \sin \theta + \overrightarrow{z} \cos \theta $ to an appropriate direction in the polar coordinates:  for the TE$_{[110]}$-mode: $\theta = 90^o$ and $\phi = 45^o$, for the TE$_{[-110]}$-mode: $\theta = 90^o$ and $\phi = 135^o$, for the TE$_{[100]}$-mode: $\theta = 90^o$ and $\phi = 0^o$, for the TE$_{[010]}$-mode: $\theta = 90^o$ and $\phi = 90^o$ and for the TM$_{[001]}$-mode: $\theta = 0^o$. Next each optical transition strength is artificially broadened by multiplication with a Gaussian distribution centred at the wavelength of the transition~\cite{Singh_2}. Finally we add all of the four Gaussian functions to calculate the total optical intensity function, \textit{f($\lambda$)}. The complete expression for the optical intensity function is given by equations 1 and 2 as follows:

\begin{widetext}
\begin{equation}
{f(\lambda)}_{T^{e_1-h_j}} = \sum_{j=1}^{4} (T^{e_1-h_j}).e^{-\frac{\displaystyle \lambda - \lambda_{e_{1}-h_{j}}}{\displaystyle (0.25)^2}}
\end{equation}
where,
\begin{equation}
T^{e_1-h_j} = (TE_{[110]}^{e_1-h_j}/TE_{[-110]}^{e_1-h_j}/TE_{[100]}^{e_1-h_j}/TE_{[010]}^{e_1-h_j}/TM_{[001]}^{e_1-h_j})
\end{equation}
\vspace{2mm}
\end{widetext}

To investigate the effect of electric fields, the potential due to a static uniform electric field is added in the diagonal of the tight-binding Hamiltonian. We consider electric fields both parallel to the growth direction [001] (labelled as $\vec{E_p}$) and anti-parallel to the growth direction (labelled as $\vec{E_a}$). The magnitude of the electric fields is varied between 0 and 45 KV/cm, which is in accordance with the range of the electric fields recently investigated in an experimental study for the design of solar cell devices~\cite{Kasamatsu_1}. The atomistic simulations are performed using nanoelectronic modeling tool NEMO-3D~\cite{Klimeck_1, Klimeck_2, Ahmed_Enc_2009}.

\begin{figure*}
\includegraphics[scale=0.28]{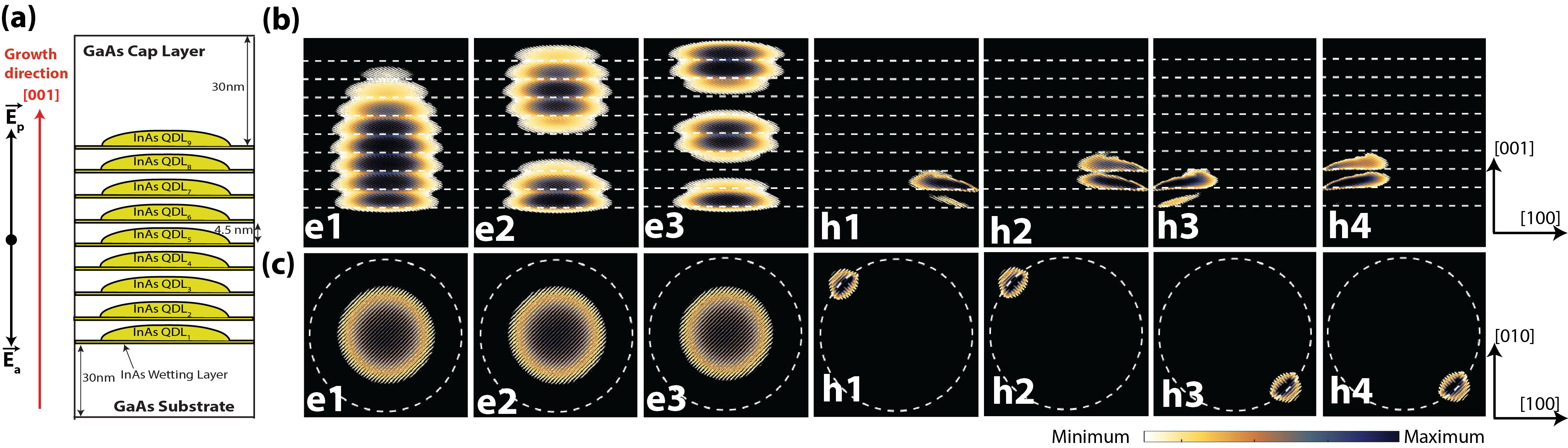}
\caption{(a) The schematic diagram of the simulated system is shown, illustrating a quantum dot molecule consisting of closely-spaced vertically-stacked nine quantum dot layers. Each QD layer is made up of a dome shaped InAs QD with the base diameter 20 nm and the height of 4 nm, residing on the top of a 0.5 nm thick InAs wetting layer. A large GaAs box surrounds the QDM, where the thicknesses of both the substrate and the cap layer are 30 nm. The lateral size of the GaAs box is 50 nm. The total number of atoms in the simulation domain is roughly 43 million. The direction of applied field is also labelled, with the field parallel to the growth axis ([001]) labelled as $\vec{E_p}$ and the field anti-parallel (opposite) to the growth axis is labelled as $\vec{E_a}$. (b) The side views of the lowest three electron states (e1, e2, e3) and the highest four hole states (h1, h2, h3, h4) are shown, indicating the spatial confinements of the wave functions. The dashed lines mark the locations of the InAs wetting layers. (c) The top views of the lowest three electron states (e1, e2, e3) and the highest four hole states (h1, h2, h3, h4) are shown, highlighting the symmetries of the wave functions. The dashed circles indicate the boundaries of the QD layers.}
\label{fig:Fig1}
\end{figure*}  

Fig.~\ref{fig:Fig1} (a) illustrates the schematic diagram of a QDM in the form of vertically-stacked closely-spaced QD layers. The QDM consists of nine QD layers (QDLs) made up of InAs material and is embedded into a large GaAs box. The geometry parameters of the QD layers (Shape = Dome, Base diameter = 20 nm, Height = 4 nm, Wetting-layer thickness = 0.5 nm, and Spacer thickness = 4.5 nm) are obtained from the recent experiments~\cite{Inoue_1, Bessho_1}, so that our theoretical results remain relevant in the context of the recently published experiments. The experimental studies~\cite{Inoue_SPIE_2010, Kasamatsu_1} have not reported any In-Ga intermixing for the QDMs. Therefore we have ignored this effect in our simulations. However, our recent studies~\cite{Usman_6, Vitto_Nano_2014} have indicated that the In-Ga intermixing have minor impact on the confinements and symmetries of the electron and hole wave functions for the single and trilayer QD systems. Consequently, we expect that the presented physical insights will remain valid in the presence of weak In-Ga intermixings.

\subsection{3 $-$ Results and Discussion}

\subsection{3.1 $-$ A brief overview of the strain effects without the application of electric fields}

Before investigating the impact of electric fields, we briefly describe the confinement of the electron and hole states in the strongly-coupled QDM (as shown by the schematic diagram of Fig.~\ref{fig:Fig1} (a)). The physical insights presented in this section will serve as a basis to understand the effects of electric fields in the later sections of this paper. The lowest three electron states (e1, e2, e3) and the highest four hole states (h1, h2, h3, h4) are plotted in Fig.~\ref{fig:Fig1}, exhibiting two different angles: (b) the side views in the [010]-plane to show the confinement of the states and (c) the top views in the [001]-plane to highlight the symmetries of the states. Here a major role is played by the underlying strain arising from the lattice mismatch of the InAs and GaAs materials, which strongly impacts both the electron and hole states. The impact of strain on the electron states is described in two ways: (1) It strongly couples the QD layers transforming the electron states of individual dots into hybridized molecular states spread over multiple QD layers $-$ for example e1 is a bonding-like state and e2 is an anti-bonding-like state and both states are spread over several QDLs. The top views of the e1, e2, and e3 states further illustrate that all of the three lowest electron states exhibit symmetric s-orbital shape. This is in stark contrast to the atomic-like states of the isolated single QD layers, where only the lowest e1 state is of s-orbital symmetry, and the first two excited (e2 and e3) states exhibit p-orbital symmetries~\cite{Usman_5}. (2) The second influence of the strain is evident from the asymmetric distribution of the electron states over the span of the QDM. For instance, the lowest electron state (e1) is quite prominently pushed towards the bottom of the QDM. Consequently, the anti-bonding-like e2 electron state is pushed towards the top of the QDM. This effect is due to the dome shapes of the QD layers, which lead to an asymmetric distribution of strain along the growth direction ([001]) and favours the bottom of the QDM for the lowest electron state.  

Contrary to the electron states, which are hybridized over several QD layers, the hole states are much more confined, even in this case of this very strongly-coupled QDM where the inter-layer separation is only 1.5 nm. This behaviour of the hole states is a direct consequence of their heavier effective mass, which reduces the impact of strain and electronic couplings. The side views of the hole states clearly indicate a restriction for the confinement of the hole states to either a single QD layer or to a couple of adjacent QD layers. Looking at the top views in Fig.~\ref{fig:Fig1}(c), another interesting property of the QDMs is observed: the hole states are confined close to the interfaces and are aligned along the [-110] direction. The confinement of the hole states close to the QD layer boundaries are related to the strong inter-dot coupling of the QD layers which implies that the QDM can be considered as a single nanostructure with a very large aspect ratio (height/base). A quantum dot with very large aspect ratio has been shown to exhibit the formation of heavy-hole pockets at the interfaces which trap the hole states~\cite{Usman_5, Narvaez_1}, therefore similar effect is observed here for the QDM. The confinement of the hole-states along the [-110] direction is related to the effect of strain, which enhances the symmetry breaking effect and makes the [110] and the [-110] directions inequivalence for the confinement of the electronic states. It has been previously shown that the strain favours the [-110] direction as the preferred alignment direction for the electronic states, leading to the lower p-state being aligned along the [-110] direction~\cite{Bester_PRB_2005, Usman_7}. Therefore all of the hole states in the QDM are aligned along this direction. 

\begin{figure*}
\includegraphics[scale=0.23]{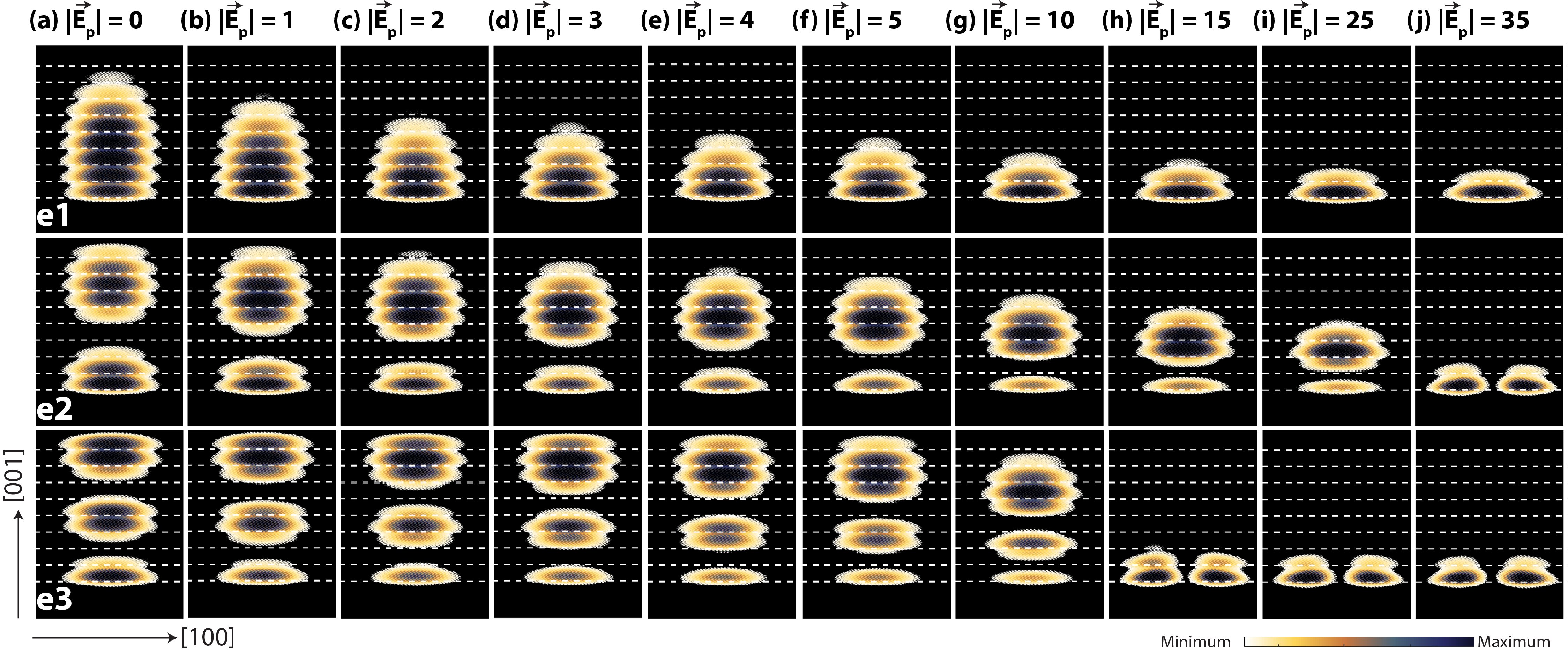}
\caption{Side views in the [010]-plane are shown for the lowest three electron states (e1, e2, e3) as a function of the electric field ($\vec{E_p}$), applied parallel to the growth direction. The magnitudes of the $\vec{E_p}$ are provided in the units of KV/cm. The dashed lines mark the locations of the InAs wetting layers. The lowest electron state e1 is rapidly pushed towards the lowest QDL. The anti-bonding state e2 maintains its character for very large values of the applied electric fields (upto 25 KV/cm). At 35 KV/cm, the coupling of the QDLs completely breaks down and the lowest three electron states roughly resemble the states of an isolated InAs QD, confined in the QDL$_1$. }
\label{fig:Fig2}
\end{figure*}
\vspace{1mm}  

\begin{figure*}
\includegraphics[scale=0.208]{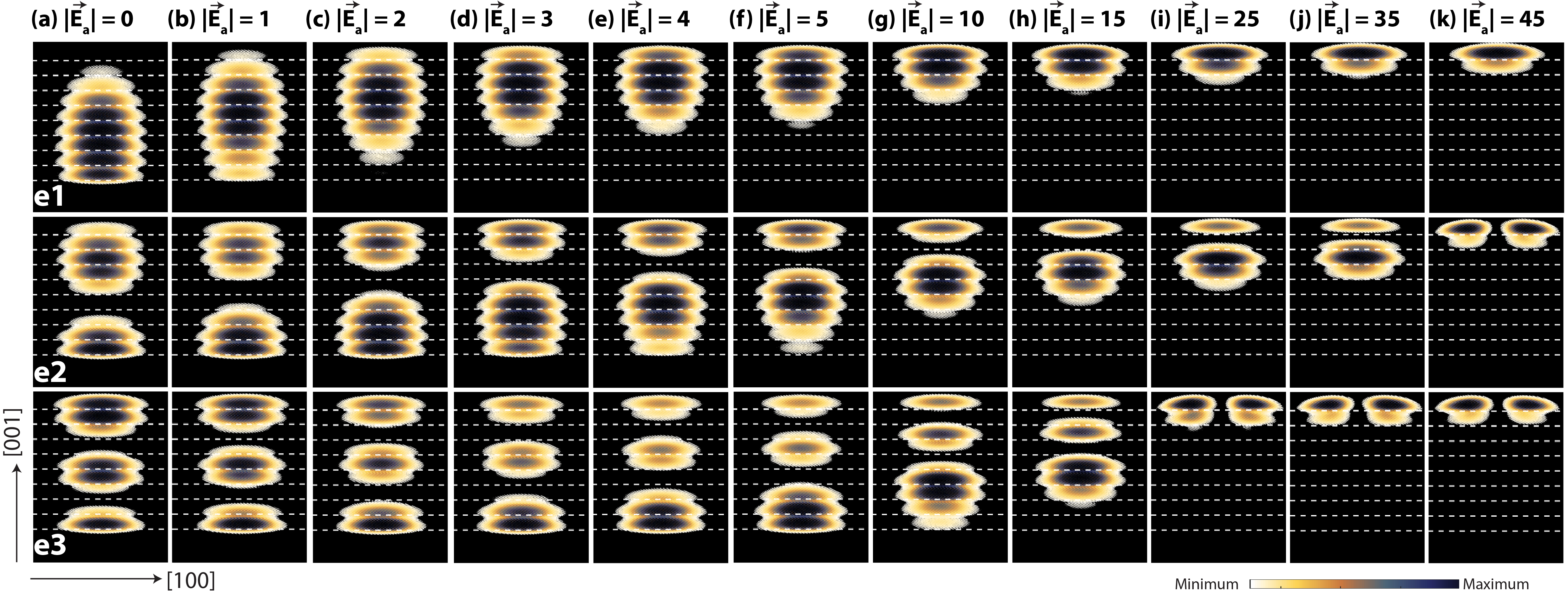}
\caption{Side views in the [010]-plane are shown for the lowest three electron states (e1, e2, e3) as a function of the electric field ($\vec{E_a}$), applied anti-parallel to the growth direction. The magnitudes of the $\vec{E_a}$ are provided in the units of KV/cm. The dashed lines mark the locations of the InAs wetting layers. The impact of electric field pushes the electrons towards the top of the QDM, opposing the impact of strain, which pushes the electron wave functions towards the bottom of the QDM. At 1 KV/cm, the effect of the field and the strain is roughly balanced and the spatial distributions of the states become symmetric with respect to the center of the QDM (QDL$_5$). At 45 KV/cm, the interdot coupling of the QDM completely breaks down and the lowest three states roughly resemble the states of an isolated InAs QD, confined in the QDL$_9$. }
\label{fig:Fig3}
\end{figure*}
\vspace{1mm}

From the aforementioned discussion, it is clear that the study and understanding of the electric field control of the electronic states in a strongly-coupled multi-layer QDM is much more involved when compared to much simpler case of single QDs. This is because in single QDs, electric field only modifies the symmetries of the electronic states and its impact on the spatial confinements is very weak -- only a small increase between electron-hole separation has been reported earlier in the literature~\cite{Fry_PRL_2000}. In contrast, due to the large volume of a QDM, the application of the electric fields can lead to much larger separations of the electron-hole states. This characteristic of the QDMs may be useful in the design of photovoltaic devices, where the reduced electron-hole recombination rates are expected to enhance charge carrier collection at the end terminals. Furthermore, the changes in the symmetries of the electronic states would also be quite different for a QDM when compared to a single QD, as all of the three electron states are s-type states and all of the hole states are interface-confined states. It should also be noted that the interplay between the strain and electric field is expected to be drastically different for the electron and hole states, as they have very different strengths of confinement in the QDM. In the next three sections, we will systematically investigate the impact of electric fields on the confined electron and hole states, separately considering their spatial confinements and symmetries.

\subsection{3.2 $-$ Electric field effect on the electron states}

In this section, we apply uniform electric fields to the QDM and calculate their impact on the lowest three electron states. As presented in the last section, the underlying strain field pushes the lowest electron state e1 towards the bottom of the QDM and the effect is in the opposite direction for the second electron state e2. Due to this inherent asymmetry of the electron states along the [001] direction, the impact of the applied electric field is also expected to be directional: a field parallel to the growth direction ($\vec{E_p}$) would support the effect of strain and a field anti-parallel to the growth direction ($\vec{E_a}$) will tend to oppose the effect of strain, cancelling its impact on the electronic structure of the QDM.  

\begin{figure}
\includegraphics[scale=0.4]{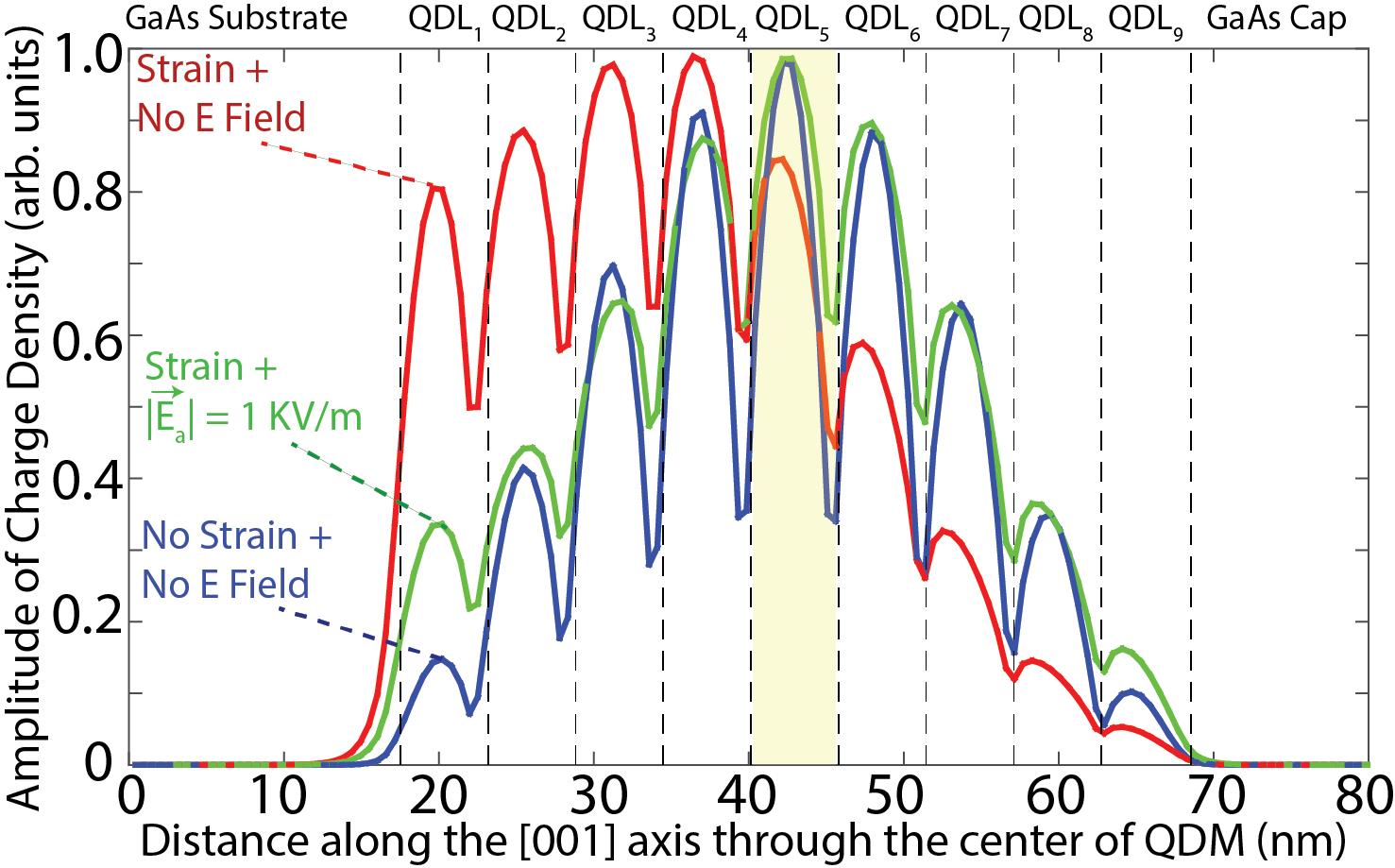}
\caption{The interplay of the strain and the applied anti-parallel electric field is highlighted by plotting one dimensional (line) cuts of the lowest electron state e1 charge density, through the center of the QDM along the growth direction for the three different cases: blue line=the effect of both strain and electric field is excluded, red line=only the effect of strain is included and the applied electric field magnitude is still zero, green line=the effect of both strain and electric field is included in the simulation. In the absence of the strain and electric field, the bonding-like state e1 is symmetrically distributed along the QDM. The strain breaks the symmetry and pushes the state towards the bottom of the QDM. A small magnitude of the electric field ($\vec{|E_a|}$ = 1 KV/cm) entirely cancels the effect of the strain and restores the symmetry of the bonding-like electron state.}
\label{fig:Fig4}
\end{figure}

Fig.~\ref{fig:Fig2} (a-j) shows the lowest three electron states as a function of the electric field magnitude ($\vec{|E_p|}$) applied along the growth direction [001], where the magnitude of the field is varied from 0 to 35 KV/cm. The electric field pushes the ground state e1 towards the bottom of the QDM, and therefore supports the effect of strain, which also pushes e1 towards the bottom of the QDM. The effect of electric field anti-parallel to the growth direction $\vec{E_a}$ is illustrated in Fig.~\ref{fig:Fig3} (a-k). In this case, the field pushes the ground state e1 towards the top of the QDM and therefore acts against the effect of the strain. As evident from Fig.~\ref{fig:Fig3} (b), the application of a small anti-parallel electric field, $\vec{|E_a|}$ = 1 KV/cm, nearly entirely cancels the effect of strain, leading to a symmetric distribution of the lowest electron state along the growth axis of the QDM. This property is further highlighted in Fig.~\ref{fig:Fig4}, where we plot one dimensional (line) cuts of the e1 charge density ($|\psi_{e1}|^2$) as a function of the distance along the [001]-axis through the center of the QDM. In the absence of both strain and electric field (shown as blue line), the charge density distribution is symmetric with its peak value at the center of the fifth (middle) QD layer of the QDM (QDL$_5$). The presence of the strain (in the absence of any electric field) slightly pushes the charge density towards the bottom of the QDM (as shown by the red line), with the peak of the charge density now residing in the fourth QD layer (QDL$_4$). The application of $\vec{|E_a|}$ = 1 KV/cm cancels the effect of strain, restoring the symmetric distribution of the wave function charge density, with its peak now again occurring in the middle (fifth) layer of the QDM (QDL$_5$).     

\vspace{1mm}

One of the most promising properties of the QDMs is the hybridization of its electron states -- the lowest two electron states e1 and e2 are bonding and anti-bonding type molecular states respectively, as shown in the Fig.~\ref{fig:Fig1} (b). This feature is a direct consequence of the strong mechanical (strain) and electronic couplings of the closely-spaced QD layers. Intriguingly this raises a question that as the electric field pushes the electron wave functions towards the edges of the QDM thereby increasing their quantum confinement, does this weaken the interdot coupling, and at what field magnitude the interdot coupling completely breaks down? A recent experimental study~\cite{Kasamatsu_1} attempted to probe this question by measuring electric field dependent PL spectra. They suggested that an electrical field of 10 KV/cm maintains the strong interdot coupling of the QD layers, which completely break down at very large field magnitudes. We theoretically address this question by analysing the spatial distributions and the character of the electron states plotted as a function of the applied electric field magnitudes in Figs.~\ref{fig:Fig2} and ~\ref{fig:Fig3}. Overall our calculations predict that the strong interdot coupling of the QDM survives large magnitudes of the applied electric fields. Even at $\vec{|E_p|}$ = 25 KV/cm and $\vec{|E_a|}$ = 35 KV/cm, the electron states maintain bonding and anti-bonding type character and are still of s-type symmetry. This supports the conclusions from the experimental data that a strong coupling of QDLs is present in the presence of moderate electric field magnitudes. Our calculations also predict that the interdot coupling completely vanishes at $\vec{|E_p|}$ = 35 KV/cm and $\vec{|E_a|}$ = 45 KV/cm fields, leading to the lowest three electron states being nearly entirely confined in the lowest (single) QD layer of the QDM. The symmetries of the states are also analogous to the states of a single isolated QD layer -- the lowest state e1 is an s-type state and the next two states (e2 and e3) are p-type states oriented along the [-110] and [110] directions, respectively. It is worth pointing out that in our calculations the difference between the $\vec{E_p}$ and $\vec{E_a}$ electric field magnitudes to achieve the breakdown of the interdot coupling is attributed to the asymmetric strain effect as discussed earlier in the section 3.1.           

\subsection{3.3 $-$ Electric field effect on the spatial confinements of the hole states}

In contrast to the electron states, which are strongly hybridized molecular states and are spread over multiple QD layers, the hole states are much more confined, residing in either single QD layers or within the adjacent couple of QD layers, as evident from the plots of the Fig.~\ref{fig:Fig1} (b). The application of the electric fields will induce an opposing force on the hole states when compared to the electron states, due to their opposite charge: $\vec{E_p}$ field will push the hole states towards the top and the $\vec{E_a}$ field will push the hole states towards the bottom of the QDM. 

\begin{figure*}
\includegraphics[scale=0.2]{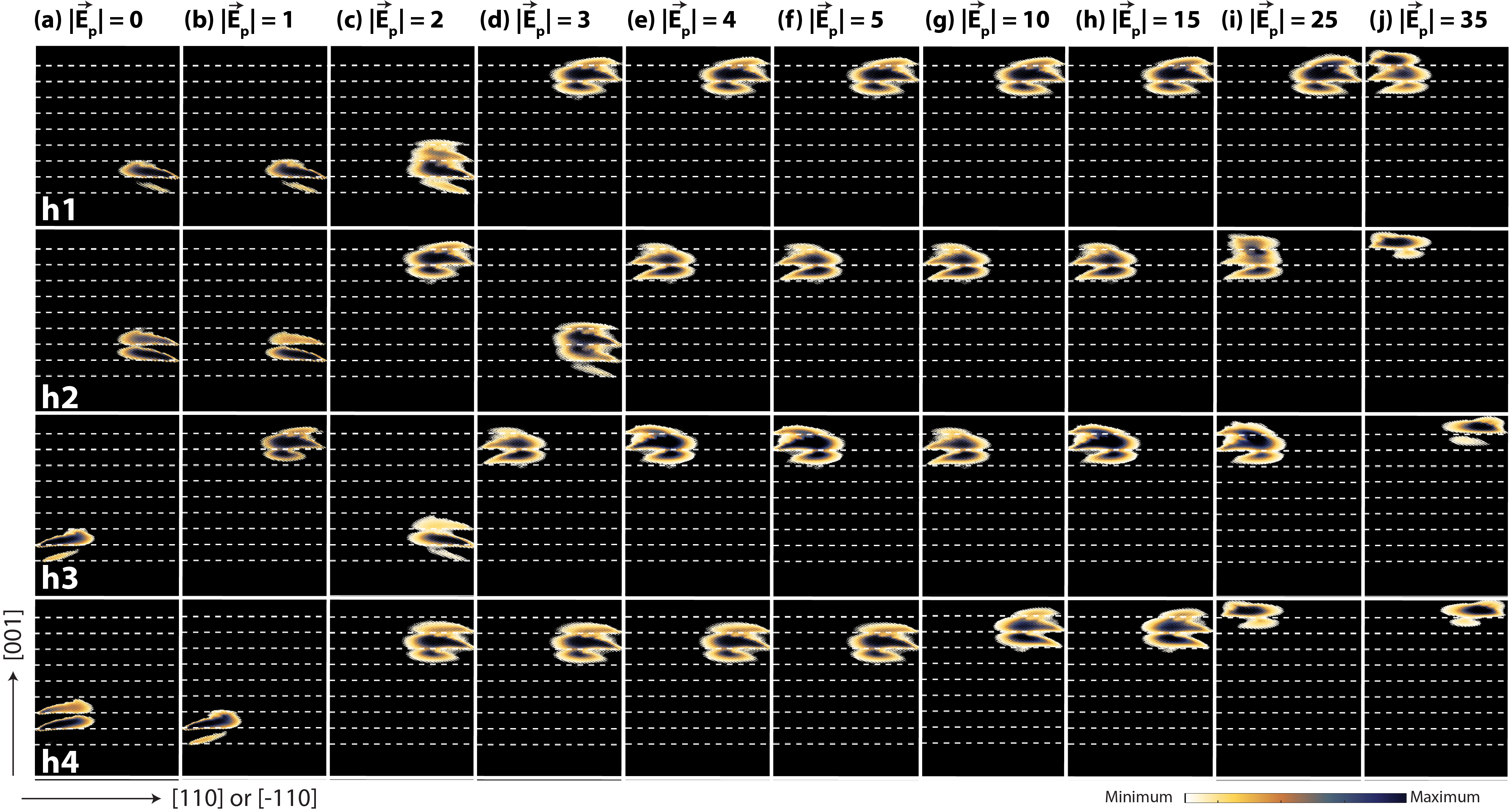}
\caption{Side views of the highest four hole states (h1, h2, h3, h4) are shown in the [010]-plane as a function of the applied electric field parallel to the growth direction ($\vec{E_p}$). The magnitudes of the $\vec{E_p}$ are provided in the units of KV/cm. The dashed lines mark the locations of the InAs wetting layers. }
\label{fig:Fig5}
\end{figure*}

Fig.~\ref{fig:Fig5} (a-j) plots the side views of the highest four hole states (h1, h2, h3, h4) in the [010]-plane, as a function of the applied electric field magnitudes in parallel to the growth direction ($\vec{E_p}$). The dotted lines indicate the position of the InAs wetting layers inside the QDM geometry. Initially without the application of any electric field (Fig.~\ref{fig:Fig5} (a)), the strain effect pushes all of the hole states towards the bottom of the QDM -- h1 is in QDL$_2$, h2 is in QDL$_2$ and QDL$_3$, h3 is in QDL$_2$, and h4 is in QDL$_2$ and QDL$_3$. Note that the h1 and h2 states are on the right edge of the QDM, and the h3 and h4 states are on the left side edge of the QDM. This arrangement is related to the symmetry of the states within the QDLs and will be discussed later in the next section when we will describe the symmetries of the hole states with the help of the top view plots. This section only focuses on the confinements of the hole states in a particular QDL along the growth direction. 

\vspace{1mm}

\begin{figure*}
\includegraphics[scale=0.2]{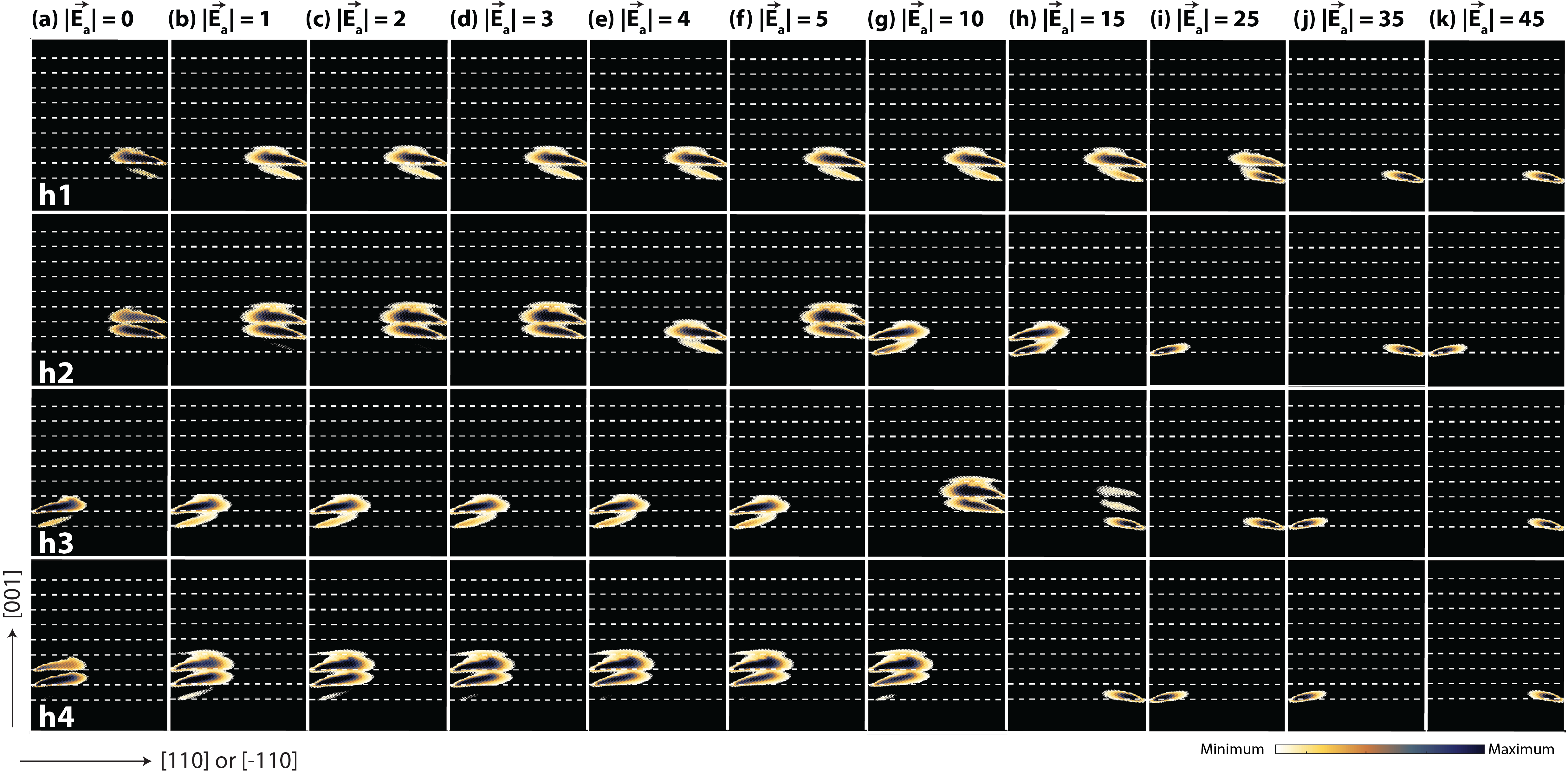}
\caption{Side views of the highest four hole states (h1, h2, h3, h4) are shown in the [010]-plane as a function of the applied electric field anti-parallel to the growth direction ($\vec{E_a}$). The magnitudes of the $\vec{E_a}$ are provided in the units of KV/cm. The dashed lines mark the locations of the InAs wetting layers.}
\label{fig:Fig6}
\end{figure*}
\vspace{1mm}

As we increase the magnitude of $\vec{E_p}$ field from 1 KV/cm to 35 KV/cm, the hole states move upward, along the growth direction of the QDM. The highest hole state h1 quickly moves to the QDL$_7$ and QDL$_8$ layers when the field is increased from 2 KV/cm to 3 KV/cm, however it only shifts inside the topmost QDL$_9$ for very large fields, $\vec{|E_p|} =$ 35 KV/cm. Our calculations also reveal that for the fields larger than 4 KV/cm, all of the highest four hole states reside close to the top of the QDM. Since the ground electron state e1 for the similar field magnitudes has moved considerably towards the bottom of the QDM: at $\vec{|E_p|}$ = 4 KV/cm, the e1 state resides in only the lowest three to four QDLs, this would considerably suppress the interband transition strengths involving e1 electron state and the hole states. We will further discuss this in the last section of this paper.     

Whilst the electric field $\vec{E_p}$ pushes the hole states towards the top of the QDM opposing the force applied by the strain fields, the field applied anti-parallel to the growth direction $\vec{E_a}$ pushes the hole states towards the bottom of the QDM, and therefore acts in support of the strain field. Fig.~\ref{fig:Fig6} (a-k) plots the side views of the highest four hole states (h1, h2, h3, h4) as a function of the applied $\vec{E_a}$ fields. The ground hole state h1, which initially resides in the QDL$_2$ at $\vec{|E_a|}$=0, quickly starts to shift into the lowest QDL$_1$ as the field magnitude increases. For 1 KV/cm $< |\vec{E_a}| <$ 15 KV/cm, we calculate a weak hybridization of the h1 state over the lowest two QDLs of the QDM. At $|\vec{E_a}|$ = 25 KV/cm, the h1 state is roughly equally present in both of the lowest QDLs (QDL$_1$ and QDL$_2$). At $|\vec{E_a}| \geq $ 35 KV/cm, the h1 state has completely moved inside the lowest QDL (QDL$_1$).  

The shift of the hole states along the QDM growth axis not only modifies the electron-hole wave function overlaps and therefore interband optical transition strengths, it also changes the heavy-hole (HH) and the light-hole (LH) character of the hole states. In the absence of electric fields, the splitting between the HH and LH states along the QDM growth direction is asymmetric~\cite{Usman_1}. The HH-LH splitting is minimum around the center of the QDM (around QDL$_5$), whereas it is much larger close to the edges of the QDM (inside the QDL$_1$ and QDL$_9$). Furthermore, due to the asymmetry of the strain along the growth direction, the HH-LH splitting is considerably larger in the bottom QDLs compared to the top QDLs of the QDM. For example, in Ref.~\onlinecite{Usman_1} the lowest layer (QDL$_1$) exhibits an HH-LH splitting of 110 meV compare to only 87 meV splitting in the top most QDL (QDL$_9$).

As we have shown that the $\vec{E_p}$ field shifts the hole states towards the bottom of the QDM, whereas the $\vec{E_a}$ field pushes the hole states towards the top of the QDM, this implies that the application of $\vec{E_p}$ field slightly increases (\textit{decreases}) the HH (\textit{LH}) character of the hole states, whereas the application of $\vec{E_a}$ field decreases (\textit{increases}) the HH (\textit{LH}) character of the hole states. Therefore the electric field response of hole states is also directional, where the HH-LH character of the hole states varies depending on the direction of the applied field. This change of the HH-LH character of the hole states is important for the polarisation-resolved optical measurements~\cite{Inoue_1}, as the TE mode optical strength is directly proportional to the HH character of the hole states, and the TM mode optical strength is directly proportional to the LH character of the hole states. Therefore, the application of an electric field to the QDM can engineer both the optical transition strengths via engineering of the electron-hole overlaps and the optical mode strengths via tuning of the HH-LH character of the hole states. 

\subsection{3.4 $-$ Electric field effect on the orientation of the hole states}

While the confinement of the hole states inside the QDM directly modifies the electron-hole wave function overlaps and therefore the interband optical transition magnitudes, the symmetry of the electron and hole states is also a critical design parameter as it strongly influences the polarisation properties of the interband optical transitions. For a single isolated QDL, it is well-known in the literature that the symmetry of the lowest electron state is s-type and it is of p-type for the next (first) two excited states~\cite{Usman_5}. It was found that the strain and the atomistic configuration of the underlying Zincblende crystals strongly impact the symmetry of the p-states~\cite{Bester_PRB_2005}. For the lattice-mismatched InAs/GaAs QDs, the influence of the strong biaxial strain splits the degeneracy of the heavy-hole and the light-hole bands, lifting the heavy-hole band higher in energy and pushing the light-hole band lower in energy~\cite{Usman_3}. Therefore the few topmost hole states have dominant heavy-hole character. This is typically true for the flat QDs with low aspect ratios (height/base $\leq$ 0.15). 

For the QDs with large aspect ratios (height/base $\geq$ 0.3), the presence of a strong impact from another effect was observed: the emergence of the heavy-hole pockets at the interfaces of the QDs~\cite{Bester_PRB_2005, Usman_7}. These heavy-hole pockets, which are formed due to a complex interaction of the strain fields and the large aspect ratios, result in the stronger confinement of the hole states closer to the interfaces of the QDs. Since the underlying strain enhances the atomistic asymmetry of the Zincblende crystals, which favours the [-110] direction in the plane of the QD, the heavy-hole pockets are deeper at the [-110] interfaces compared to the other in-plane directions. This leads to the highest few hole states being oriented along the [-110] direction. The large QDMs, consisting of very closely-spaced and strongly-coupled QDLs such as being studied in this work, could be considered as QDs with very large aspect ratios. Therefore they exhibit similar behaviour: the highest few hole states being confined at the [-110] interfaces of the QDLs inside the pockets of the hole band edges. The plots of the top views of the highest four hole states in the [001]-plane are shown in Fig.~\ref{fig:Fig1} (c) and confirm this notion.   

\begin{figure*}
\includegraphics[scale=0.225]{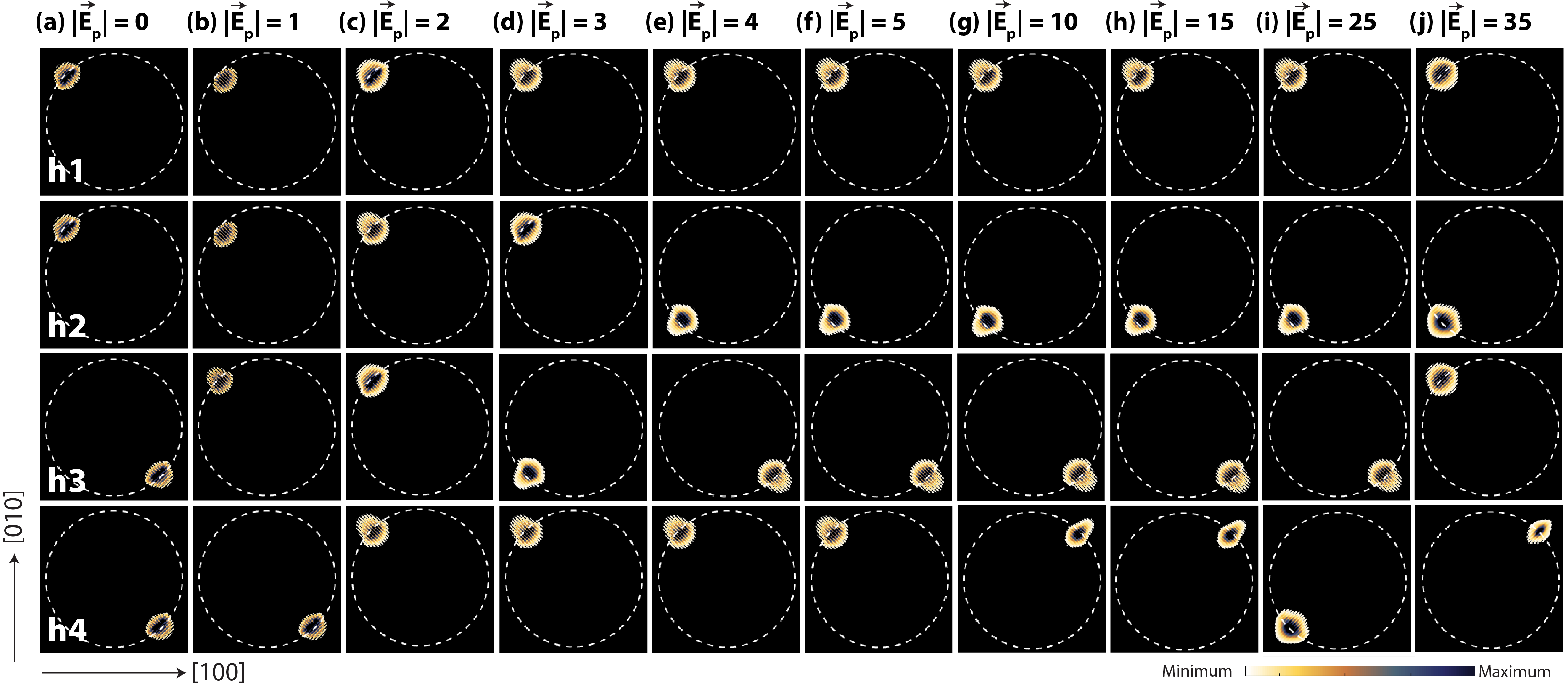}
\caption{Top views of the highest four hole states (h1, h2, h3, h4) are shown in the [001]-plane as a function of the applied electric field parallel to the growth direction ($\vec{E_p}$). The magnitudes of the $\vec{E_p}$ field are provided in the units of KV/cm. The dashed circles mark the boundaries of the InAs QD regions. }
\label{fig:Fig7}
\end{figure*}
\vspace{1mm}

\begin{figure*}
\includegraphics[scale=0.2]{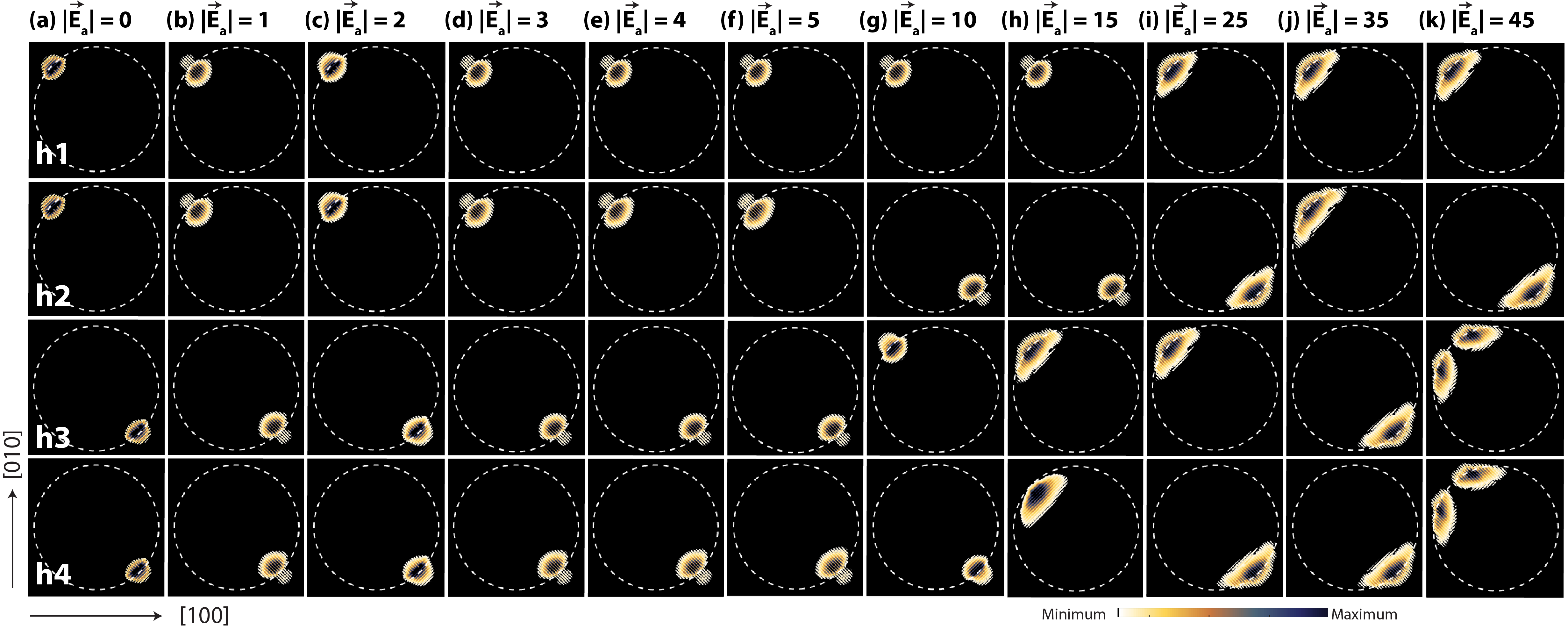}
\caption{Top view of the highest four hole states (h1, h2, h3, h4) are shown in the [001]-plane as a function of the applied electric field anti-parallel to the growth direction ($\vec{E_a}$). The magnitudes of the $\vec{E_a}$ are provided in the units of KV/cm. The dashed circles mark the boundaries of the InAs QD regions.}
\label{fig:Fig8}
\end{figure*}
\vspace{1mm}   

One would naively expect that the application of the electric fields in parallel or anti-parallel to the growth direction should not impact the orientation of the hole states in the plane of the QDLs, as it should primarily depend on the in-plane fields. However, as discussed in the last section, that the application of $\vec{E_p}$ and $\vec{E_a}$ fields results in the tunnelling of the hole wave functions from one QDL to the other QDL. Consequently the orientation of the hole wave functions also depend on the nature of the hole pockets belonging to that particular QDL in which a hole state resides at an applied field magnitude. Therefore changes in the orientation of the hole states could be observed as a function of the electric fields. 

To investigate the effect of electric fields on the orientations (symmetries) of the hole states, we vary $\vec{E_p}$ and $\vec{E_a}$ fields and plot the top views of the field-dependent hole wave function in the [001]-plane as shown in Figs.~\ref{fig:Fig7} (a-j) and ~\ref{fig:Fig8} (a-k). The dotted circles roughly mark the boundaries of the QDLs. The topmost row in Fig.~\ref{fig:Fig7} shows that the application of an electric field in parallel to the growth direction, $\vec{E_p}$, does not impact the orientation of the ground hole state h1. Therefore the h1 state remains oriented along the [-110] direction for all of the applied field magnitudes. Interestingly the other three hole states h2, h3, and h4 exhibit some changes in their orientations when the magnitude of the $\vec{E_p}$ field is increased. For example, the h2 state is oriented along the [-110] direction for $\vec{|E_p|} \leq$ 3 KV/cm and is oriented along the [110] direction for $\vec{|E_p|} >$ 3 KV/cm. Even a stronger variation in the orientations of the h3 and h4 states is observed. At 35 KV/cm, the interdot coupling breaks down and the electron and hole states are confined in the individual QDLs. Therefore, the orientation of the hole states for 35 KV/cm is what we expect for a single QD: the hole states are alternatively oriented along the [-110] and the [110] directions~\cite{Usman_2}. This implies that the in-plane polarisation modes TE$_{[110]}$ and TE$_{[-110]}$ will be of equal magnitudes, despite the confinements of the hole states being strongly directional, as previously reported for the single QDs~\cite{Usman_2}.     

For the highest hole state h1, the behaviour is same for the field applied anti-parallel to the growth direction, $\vec{E_a}$: the h1 state is oriented along the [-110] direction independent of the applied field magnitudes. Therefore we conclude that for the h1 state, the orientation is independent of both, the field magnitude and the field direction. Consequently if the optical transition is designed to involve only the h1 state, the only dominant optical mode would be the TE$_{[-110]}$ mode. The h2 state exhibits a slightly different behaviour under the influence of $\vec{E_a}$ field, when compared to the $\vec{E_p}$ field: the h2 state is also oriented along the [-110] direction for all of the applied $\vec{E_a}$ magnitudes. However, it switches confinement in the HH pockets from upper-left corner of the QDL to the lower-right corner for the certain values of the fields as shown in the second row of the Fig.~\ref{fig:Fig8}. Finally we also calculate that the h3 and h4 states also exhibit slightly different $\vec{E_a}$ field response, compared to the $\vec{E_p}$ field. One important difference is the emergence of p-type hole h3 and h4 states at $\vec{|E_a|}$ = 45 KV/cm field. 

Overall, from the examination of the field dependent hole state orientations, we conclude this section by stating that the fields applied along the perpendicular directions to the growth direction modify the symmetries of the hole states in the plane of the QDM. This unintuitive result is attributed to the complex behaviour of the HH pockets found at the interfaces of the QDLs inside the QDM. 

\subsection{3.5 $-$ Transition Energies and Band-gap wavelengths}

\begin{figure*}
\includegraphics[scale=0.26]{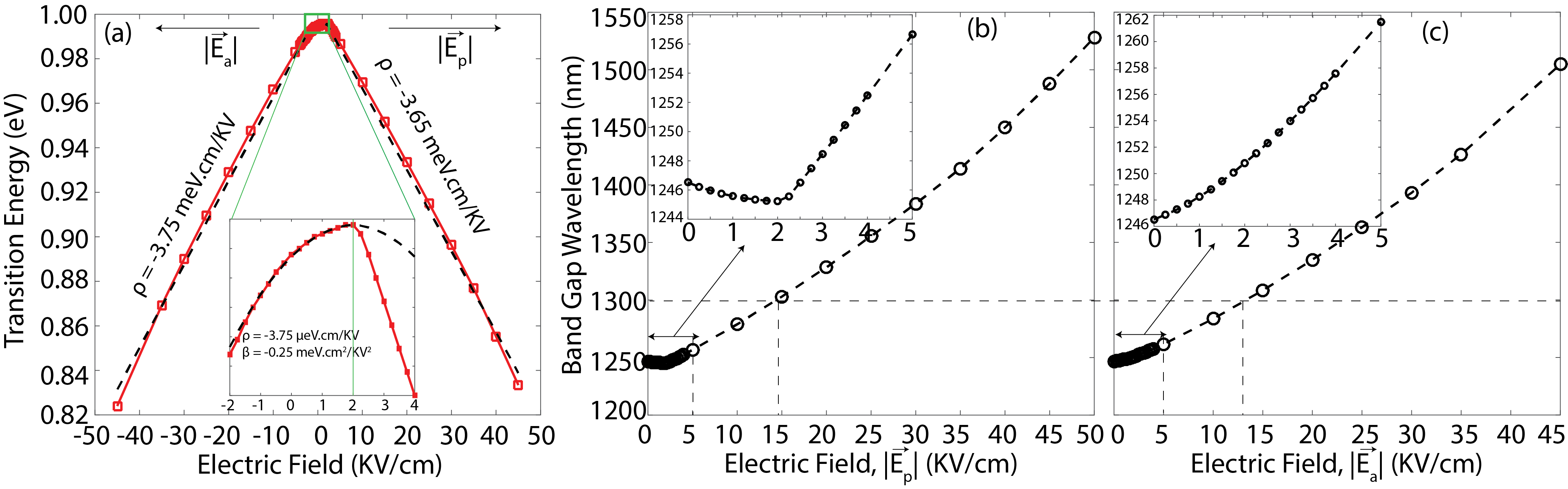}
\caption{(a) Transition energies (e1-h1) are plotted as a function of the magnitudes of the applied electric fields. For the large electric field magnitudes $|\vec{E}_{p/a}| >$ 2 KV/cm, the field dependence of the transition energy is nearly linear indicating negligible polarizibility ($\beta$). For $|\vec{E}_{p/a}| \leq$ 2 KV/cm, the transition energy exhibits a perfect parabolic dependence on the electric field implying very small dipole moment ($\rho$) (see the inset). (b, c) Band-gap wavelength (1240/e1-h1) in nm is plotted as a function of the magnitudes of the applied electric fields: (b) the field direction is parallel to the growth direction ($\vec{E_p}$) and (c) the field direction is anti-parallel to the growth direction ($\vec{E_a}$). The inset plots highlight the wavelength dependence for the small magnitudes of the electric fields. }
\label{fig:Fig9}
\end{figure*}

The electric field dependence of the ground state transition energy ($Trans_e$=e1-h1) is plotted in Fig.~\ref{fig:Fig9} (a), where the positive and negative values of the electric field indicate the $\vec{E_p}$ and $\vec{E_a}$ fields, respectively. For a QD system, the $Trans_e$ value is represented by the following equation~\cite{Usman_IWCE_2009}:

\begin{equation}
Trans_e( |\vec{E}_{p/a}| ) = Trans_e( 0 ) + \rho |\vec{E}_{p/a}| + \beta |\vec{E}_{p/a}|^2
\end{equation}
\noindent
where $Trans_e( |\vec{E}_{p/a}| )$ is the transition energy value at $|\vec{E}_{p/a}|$ field,  $Trans_e$(0) is the transition energy at reference field (here we assume it to be zero), $\rho$ depends on the built in dipole moment due to the electron-hole separation induced by the applied electric fields, and $\beta$ is a measure of the polarizibility of the electron and hole wave functions. For a typical single InAs QD, the $Trans_e$ dependence on the electric field is parabolic~\cite{Usman_IWCE_2009, Vogel_2007}, indicating a large magnitude for the polarizibility $\beta$ and a very small magnitude for the dipole moment $\rho$ -- a direct consequence of the small space available for the electron and hole separation. For QD stacks, made up of a large number of QD layers, electric fields are capable of pushing electron and hole wave functions far away from each other (even at the opposite ends of the stack for large fields), and therefore can lead to large separations between the electron and hole wave functions. This results in much larger values for the dipole moment $\rho$ compared to the polarizilibilites $\beta$ -- hence the field response of the QD stacks is characterized by a linear dependence of the $Trans_e$ on the applied electric fields. 

Fig.~\ref{fig:Fig9} (a) overall demonstrates a linear dependence of the $Trans_e$ value on the applied field magnitudes for the studied QDM. For the large values of the applied electric fields, $|\vec{E}_{p/a}| >$ 2 KV/cm, the electric field dependence of the $Trans_e$ is nearly perfectly linear and therefore can be represented by the following equation:

\begin{equation}
Trans_e( |\vec{E}_{p/a}| ) = Trans_e( 0 ) + \rho |\vec{E}_{p/a}|
\end{equation}  
\noindent
The black dotted lines in the Fig.~\ref{fig:Fig9} (a) plots the fits from the equation 4, with the values of $\rho$ used as -3.65 meV.cm/KV and -3.75 meV.cm/KV for the positive and the negative electric fields, respectively. These values of the $\rho$ are significantly larger compared to the single QD values -- several orders of magnitude smaller values have been reported in the literature for the single QDs~\cite{Usman_IWCE_2009}.  

It is also noted that for the small values of the applied electric fields, $|\vec{E}_{p/a}| \leq$ 2 KV/cm, the $Trans_e$ values exhibit a parabolic dependence on the electric field magnitudes. This is clear from the inset plot in Fig.~\ref{fig:Fig9} (a), where the dotted black line shows the fitting from the equation 3 with the values of $\rho$ = -3.75 $\mu$eV.cm/KV and $\beta$ = -0.25 meV.cm$^2$/KV$^2$.  While the magnitude of $\rho$ is decreased by three orders of magnitude, the value of the polarizibility $\beta$ is significantly large. This is also considerably larger compared to the measured values of -2 to -5 $\mu$eV.cm$^2$/KV$^2$ for the single QDs~\cite{Vogel_2007}. Therefore we conclude that a strongly coupled QDM can be operated in both regimes: a large polarizibility (\textit{dipole moment}) regime for the small (\textit{large}) magnitudes of the applied electric fields.

It should also be noted from the inset of the Fig.~\ref{fig:Fig9} (a) that the maximum value of the $Trans_e$ parameter occurs at 2 KV/cm electric field, rather than 0. This slight asymmetry is a direct consequence of the underlying strain field that pushes the electron state e1 towards the top of the QDM and the hole state h1 towards the bottom of the QDM as discussed earlier in the section 3.1 (see also Fig.~\ref{fig:Fig1} for reference). 

Another parameter of interest for the optoelectronic devices is the band-gap wavelength. Typically large strain fields blue shift the optical wavelengths of the QD systems and therefore make them unsuitable to operate at the optical telecommunication wavelengths (1.3 $\mu$m and 1.55 $\mu$m). It has been a great challenge to achieve the optical emissions from the QD systems at 1.3 $\mu$m and 1.55 $\mu$m wavelengths, where several different techniques have been used in the literature to overcome this problem. The most widely used QD system offering the long wavelength emission is QD inside a quantum well structure, where strain engineering results in the red shift of the PL peak~\cite{Usman_2, Usman_3}. The application of the electric fields also induces large shifts in the electron and hole confinement energies, which is commonly known as the quantum confined Stark effect (QCSE). The QCSE results in reducing the band-gap energy of the QD system and therefore red shifts the band-gap wavelengths. Here we investigate the band-gap wavelength increase as a function of the applied electric fields. For a single QD, the QCSE is only dependent on the bending of the band edges and therefore it primarily exhibits a linear dependence on the magnitude of the electric field~\cite{Fry_PRL_2000}. Similar behaviour has been observed for weakly-coupled bilayer QD systems~\cite{Usman_4}. However for a strongly-coupled QDM, such as the one being investigated in this study, the strong interdot couplings lead to a non-trivial dependence of the QCSE on the magnitude of the electric fields. We have previously discussed in the section 3.1 that the electron states are strongly coupled hybridized states whereas the hole states are much more confined and typically present only in one or two QDLs of the QDM. This implies that the electric field would lead to the tunnelling of the hole states between the QDLs and therefore adds an extra level of complexity in the QCSE behaviour.  

\vspace{1mm}

We plot the band-gap wavelengths for the QDM as a function of both $\vec{E_p}$ and $\vec{E_a}$ electric field magnitudes in Figs.~\ref{fig:Fig9} (b) and (c), respectively. The band-gap wavelengths are calculated by converting the energy difference of the lowest electron state e1 and the highest hole state h1 to a corresponding wavelength on the nm scale: (1240/e1-h1). Since the strain distribution along the growth direction is asymmetric, the electric field induced QCSE in the band-gap wavelengths is also to be different for the $\vec{E_p}$ and $\vec{E_a}$ fields. The difference mainly arises from the tunnelling of the hole states as a function of the electric fields. For the $\vec{E_p}$ field, the hole state h1 shifts from the bottom to the top of the QDM and therefore experiences a strong interdot tunnelling effect. This is reflected in the inset of Fig.~\ref{fig:Fig9} (a), where a strong deviation from the typical linear increase of the band-gap wavelength is calculated. However this effect only lasts for very small magnitudes of the $\vec{E_p}$ field. At 2 KV/cm field, the hole state h1 has already moved to the top of the QDM, so the tunnelling related effect disappears. Further increase in the $\vec{E_p}$ field magnitude leads to roughly linear increase in the band-gap energy.

The application of the field anti-parallel to the growth direction, $\vec{E_a}$, does not show much deviation from the linear behaviour, even for the small magnitudes of the field. This is because the h1 state is already close to the base of the QDM at zero field, and therefore it only experiences a weak tunnelling effect when $\vec{|E_a|}$ is switched on. The inset of the Fig.~\ref{fig:Fig9} (b) demonstrates this behaviour for  $\vec{|E_a|} \leq$ 2 KV/cm.  

\subsection{3.6 $-$ Interband optical transition strengths}

For a variety of optoelectronic devices such as semiconductor optical amplifiers (SOAs), polarisation-resolved interband optical transitions are a critical design component. Several experimental~\cite{Inoue_1, Harada_1} and theoretical~\cite{Usman_1, Usman_9} studies have explored this parameter to attain isotropic polarisation response, where the optical mode strengths are independent of the polarisation direction. Polarisation-dependent interband optical transition strengths (TE and TM modes) depend on two factors: (1) the spatial overlap between the electron and the hole wave functions and (2) the heavy-hole (HH) and the light-hole (LH) character of the hole states. In QDMs, the strong coupling between the QDLs lead to enhanced LH character of the valence band states. Furthermore the spatial overlap between the electron and hole states is decreased due to the interfacial confinements of the hole states (see Fig.~\ref{fig:Fig1 (b)} for reference). The two factors lead to strong modifications of the TE and TM polarisation modes. In particular, the previous experimental measurements~\cite{Inoue_1} and theoretical calculations~\cite{Usman_1} have revealed a unique property for the QDMs, where the TM mode strength becomes comparable to the TE mode strength -- a characteristic critically important for the design of semiconductor optical amplifiers (SOAs). This property is not accessible from single QDs, where the TE mode is always dominant and the TM mode is very weak, resulting in a strong anisotropy of the polarized absorption spectra~\cite{Usman_2}.    

\begin{figure*}
\includegraphics[scale=0.3]{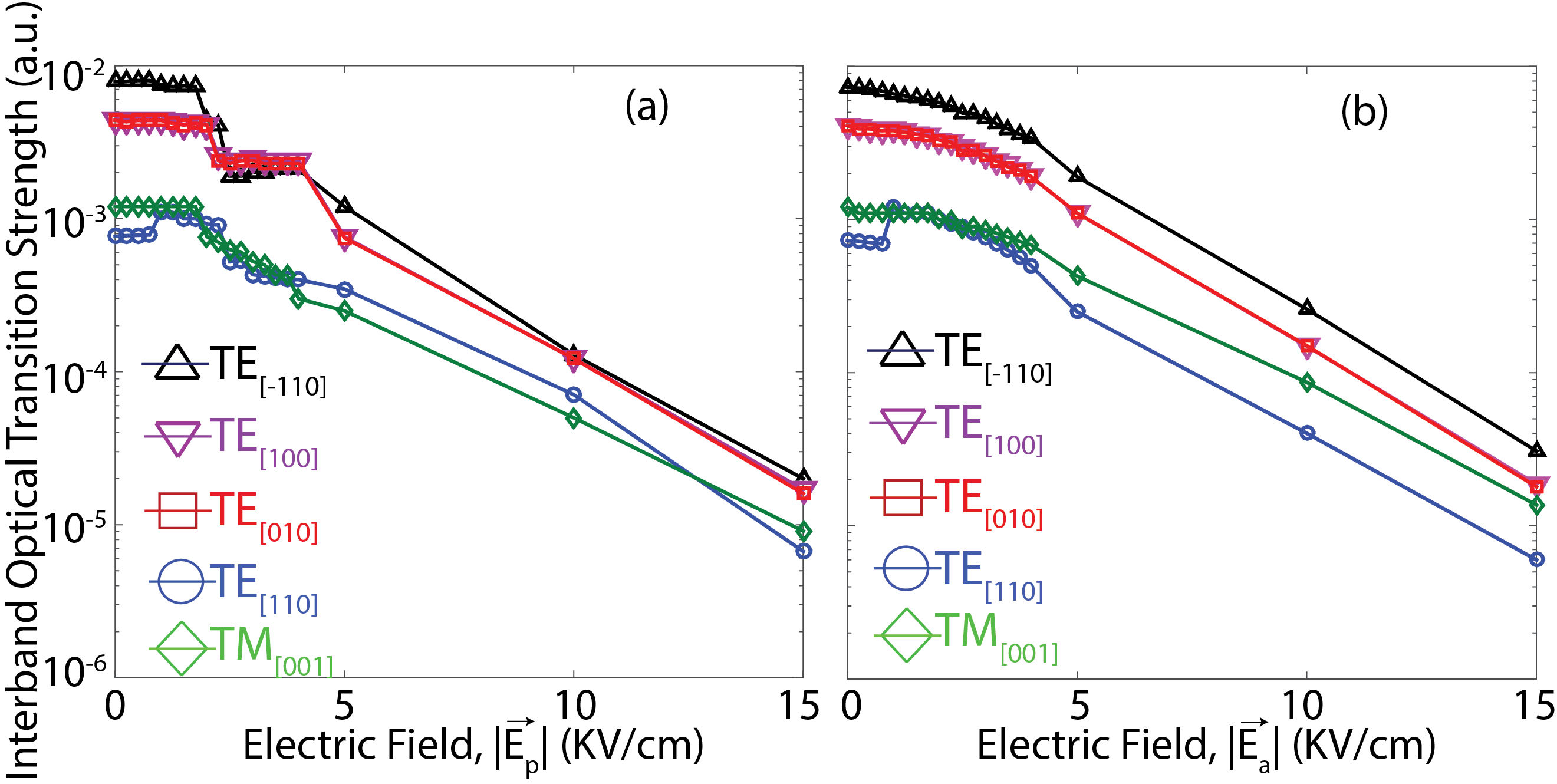}
\caption{Interband optical transition modes (TE and TM modes) calculated form the equations 1 and 2 are plotted as a function of the applied electric fields: (a) The applied field direction is parallel to the growth direction of the QDM and (b) the applied field direction is anti-parallel to the growth direction of the QDM.}
\label{fig:Fig10}
\end{figure*}

In this section, we examine the polarisation resolved interband transition modes as a function of the applied electric fields. Since the electric field has been shown to strongly modify the electron and hole spatial confinements, the optical mode strengths are also expected to exhibit correspondingly large variations. For a complete spectra of the optical modes, we investigate the TE modes in the four (high symmetry) in-plane directions, which are of particular relevance to the experimentalist: [-110], [110], [100], and [010]. The corresponding TE modes are labelled with the polarisation directions in the subscripts. The TM mode is always calculated along the growth [001] direction of the QDM. Fig.~\ref{fig:Fig10} plots the polarisation mode strengths as a function of the applied electric fields: (a) when the field is parallel to the growth direction, $\vec{|E_p|}$ and (b) when the field is anti-parallel to the growth direction, $\vec{|E_a|}$. 

\vspace{1mm}

The application of the electric fields shifts the electron and hole wave functions in the opposite directions and therefore reduces the electron-hole wave function overlaps. Therefore, overall all of the optical mode strengths decrease as a function of the electric field. A careful analysis of the plots of Fig.~\ref{fig:Fig10} highlights the following insights: (1) The decease in the optical mode strengths for small magnitudes of the electric fields is smoother for the $\vec{E_a}$ field, when compared to the $\vec{E_p}$ field. This is simply due to the fact that the hole wave functions experience significantly weaker interdot tunnelling under the application of the $\vec{E_a}$ field, compared to the $\vec{E_p}$ field. (2) The TE$_{[-110]}$, TE$_{[100]}$, and TE$_{[010]}$ optical modes remain much stronger than the TM$_{[001]}$ mode for all magnitudes of the applied fields, and it is also independent of the field direction. Therefore, isotropic polarisation is not accessible in these polarisation directions with or without fields. (3) The magnitudes of the TE$_{[100]}$ and TE$_{[010]}$ modes are always same, independent of the electric field magnitudes and directions. This is due to the fact that the electric fields in this study are only applied along the growth [001] direction and therefore do not have any directional effect for the in-plane polarisation modes. Furthermore, the hole wave functions are confined along the [110] or [-110] directions, therefore the [100] and the [010] directions are equivalent with respect to the electron and hole wave function overlaps. (4) At the zero field value, TE$_{[110]} >$ TM$_{[001]}$. However for small magnitudes of the electric fields (for both $\vec{E_p}$ and $\vec{E_a}$), approximately varying from 2 KV/cm to 4 KV/cm, the TE$_{[110]}$ mode strength is roughly equal to the TM$_{[001]}$ mode strength. This implies that the application of the small electric fields can be used to realize isotropic polarisation from QDMs. (5) Due to the large separation between the electron and hole wave functions, the optical mode strengths rapidly decrease as a function of the applied electric fields. At 15 KV/cm, we calculate that the optical mode strengths have been reduced by approximately two to three orders of magnitudes. Therefore, very weak interband optical absorption is expected at the large electric fields. This is expected to be important for the design of photovoltaic devices such as intermediate-band solar cells where optically generated electron-hole pairs will be prohibited from reabsorption via interband transitions.   

\section{4 $-$ Conclusions}

In conclusion, we have provided a comprehensive understanding of the electric field control of the electronic and optical properties of the multi-layer strongly-coupled quantum dot molecules (QDM). Electric fields are applied in both parallel ($\vec{E_p}$) and anti-parallel ($\vec{E_a}$) directions with respect to the growth axis ([001]) of the QDM. It is found that the asymmetric effect of the strain on the electronic wave functions can be entirely balanced by the application of a small electric field of the order of 1 KV/cm along a direction anti-parallel to the growth direction. We have provided a detailed investigation of the electron and hole wave functions under the influence of electric fields by plotting the top and side views, which would serve as a useful guide for the experimentalists working on the devices based on the QDMs. Our calculations highlight a complex interplay between the applied electric fields and the strong interdot couplings, which control the spatial confinements of the electron and hole wave functions in different QDLs of the QDM. It is revealed that the interdot couplings survive large electric field magnitudes and a complete break down occurs at $|\vec{E_p}| >$ 35 KV/cm or $|\vec{E_a}| <$ 45 KV/cm. At these large electric fields, the electronic states reduce to single quantum dot states confined at the opposite ends of the QDM. This mimics a transformation from a type-I band structure to a type-II band structure, which is a requirement for the design of intermediate-band solar cell devices. By analysing the ground state transition energy data, we highlight an interesting property of the QDMs that it could be operated in two regimes: for the small electric field magnitudes, the QDM exhibits large polarizibilities and negligible dipole moment strengths; in contrast for the large electric field magnitudes, the QDM demonstrates negligible polarizibilities and the field response is dominant by the dipole moment. The band-gap wavelength of the QDM red shifts as a function of the applied electric fields due to the quantum confined Stark effect (QCSE) and increases to 1.3 $\mu$m at the 15 KV/cm field. However, the reduced overlap of the electron-hole wave functions results in the optical transition strengths reduced by two to three orders of magnitude, thereby a much smaller optical gain/oscillator strength is expected under the influence of the electric fields. We have also provided a detailed investigation of the polarisation-resolved interband optical transition modes (TE and TM modes) as a function of the applied electric fields. Overall the polarisation mode strengths decrease as a function of the applied electric fields due to the reduced electron-hole wave function overlaps. However more importantly, the application of the small electric fields is found to be helpful to realize the isotropic polarisation response from the QDMs, with TE$_{[110]} \approx$ TM$_{[001]}$. This condition is crucial for the design of semiconductor optical amplifiers (SOAs) where the isotropic polarization is highly desirable. Overall, our multi-million atom calculations provide a critical and reliable theoretical guidance for the experimentalists on a topic where little theoretical knowledge is available in the existing literature. The physical insights presented will be highly useful to engineer electronic and optical properties of the QDMs for the design of a variety of optoelectronic, photovoltaic, and quantum information science devices.

\textbf{\textit{Acknowledgements:}} Computational resources are acknowledged from National Science Foundation (NSF) funded Network for Computational Nanotechnology (NCN) through \url{http://nanohub.org}. Parts of this work were done at the Tyndall National Institute, Lee Maltings, Dyke Parade, Cork Ireland.

\bibliographystyle{apsrev4-1}

%

\end{document}